\documentclass[twocolumn, prl]{revtex4}
\usepackage{graphicx}
\usepackage{epsfig}
\begin{document}
\title{Squeezing Out the Entropy of Fermions in Optical Lattices}
\author{Tin-Lun Ho and Qi Zhou}
\affiliation{Department of Physics, The Ohio State University, Columbus, OH 43210}
\begin{abstract}
 At present, there is considerable interest in using atomic fermions in optical lattices to emulate the mathematical models that have been used to study strongly correlated electronic systems. Some of these models, such as the two dimensional fermion Hubbard model, are notoriously difficult to solve, and their key properties remain controversial despite decades of studies.  It is hoped that the emulation experiments will shed light on some of these long standing problems. A successful emulation, however, requires reaching temperatures as low as $10^{-12}$K and beyond, with 
entropy per particle far lower than what can be achieved today.  
Achieving such low entropy states is an essential step and a grand challenge of the whole emulation enterprise. In this paper, we point out a method to literally squeeze the entropy out from a Fermi gas into a surrounding Bose-Einstein condensed gas (BEC), which acts as a heat reservoir. This method allows one to reduce the entropy per particle of a lattice Fermi gas to a few percent of the lowest value obtainable today.
\end{abstract}

\maketitle

Currently, many laboratories are trying to realize the anti-ferromagnetic (AF) phase of the 3D Hubbard model using ultra-cold fermions in optical lattices\cite{Cho}.  
Although the AF phase  is well known in condensed matter, its realization will be a major step for the emulation  program since it requires overcoming the serious challenge mentioned above which  is common to all cold atom emulations\cite{HoZhou}. To understand the origin of the problem, recall that the strongly correlated states of lattice fermions emerge in the lowest Bloch band of the optical lattice. To put all the fermions in the lowest band, a sufficiently deep optical lattice is required. 
In such deep lattices, many known methods of cooling fail.  
For example, standard evaporative cooling does not work because the magnetic repulsive potential used in the evaporation process is not strong enough to overcome the deep lattice. 
%For example, standard evaporative cooling ceases to work because the usual magnetic de-confining potentials with which it ejects particles are not strong enough to overcome the deep lattice. 
As a result, all current experiments on lattice quantum gases resort to the conventional cooling scheme, first cooling the quantum gas in a harmonic trap (without optical lattice) to the lowest entropy state possible, and then turning on the lattice adiabatically\cite{adiabatic}\cite{Kohl}. The hope is that one could reach the state of interest when the lattice depth is sufficiently high.   

To realize these strongly correlated states with the current scheme, it is necessary that the entropy of the gas prior to switching on the lattice be less than that of the strongly correlated state one wishes to achieve.
%For this scheme to work, the entropy of the desired strongly correlated state must exceed what can be achieved using the current method to cool a quantum gas in a harmonic trap. 
%it is necessary that the desired strongly correlated state has an entropy high enough that can be reached by a quantum gas cooled in a harmonic trap. 
For 3D fermion Hubbard models, recent studies\cite{AG}  show that it is possible to reach the antiferromagnetic (AF) phase slightly below the Neel temperature $T_{N}$ with the conventional cooling method. 
A similar calculation by Tremblay et.al.\cite{Tremblay} for the 2D Hubbard model, however, showed that the conventional scheme cannot reach even the pseudo-gap regime, which exists at a higher 
temperature than the anticipated superconducting phase. 
Although these calculations are for homogenous systems, they apply to confining traps as long as the majority of the sample is a Mott insulator with one fermion per site. These studies show that even under optimal conditions, one can at most reach the AF phase close to the magnetic ordering, but not low enough temperatures to study ground state properties. 

The problem of the conventional method is that it can only cool atoms before they are loaded onto an optical lattice. There is no way to reduce the entropy further once the lattice is switched on. 
Thus, the lowest entropy  attainable today in an optical lattice plus harmonic trap is just the lowest entropy achievable in a harmonic trap alone. 
% all its cooling power exist only  before loading the atoms onto the lattice.  There is no way to reduce the entropy further once the lattice is switched on. 
%One is therefore stuck with the lowest entropy per particle attainable today in a single trap. 
In the case of Fermi gases, it is $S/N = \pi^2(T/T_{F})$\cite{Castin}, where $T_{F}$ is the Fermi temperature in the trap.  For  $T/T_{F}= 0.05$, which is very much the limit today, we have $S/N \sim 0.5$. 
Any strongly correlated states with lower entropy are unreachable by this method. 
It is therefore important to find ways to reduce the entropy of the system significantly below this value. 

\section{(A) Our  entropy removal scheme:} 
The purpose of this paper is to present a scheme to produce a lattice Fermi gas with about one fermion per site with entropy per particle much lower than what can be achieved today.  Our method is based on the principle of entropy {\em redistribution} and the removal of entropy by {\em isothermal compression}.  It consists of the following steps:

\noindent ${\bf (I)}$ We immerse the lattice fermions in a BEC which acts as a heat reservoir.  The fermions are confined in a harmonic trap and a strong optical lattice.  The bosons sees a much weaker trapping lattice than the one that traps for the fermions.  They are confined in a loose trap and cover the entire fermion system.   The traps for bosons  and fermions are species specific\cite{specific} so that they can be varied separately.  (See below for discussions of these potentials.)
 
\noindent 
{\bf (II)} We compress the fermion harmonic trap adiabatically to turn the fermions at the center into a band insulator, which has two fermions per site and essentially zero entropy. During this process, a substantial amount of the original fermion entropy is pushed into the bosons, while the entire system has little temperature change because of the large heat capacity of the BEC compared to the lattice fermions. 
Hence, even though the process is adiabatic for the entire Bose-Fermi system, it is essentially isothermal as far as  the fermions are concerned. 

\noindent {\bf (III)} After pushing out the fermion entropy into the BEC, we remove it by evaporating away the bosons all at once, leaving the remaining fermions to equilibrate. 
Since the band insulator is incompressible, only its density and entropy near the surface are affected during this process. This in turn severely limits entropy re-generation during the equilibration process. 
As a result, the entropy of the re-thermalized band insulator has a similar ultra-low value as before  boson evaporation. 

\noindent {\bf (IV)} We open up the the fermion harmonic trap adiabatically to lower the density of the lattice fermions. In this way, one can produce a Mott insulator or other states with fractional filling with the same ultra-low entropy. 

Before proceeding, we return to discuss the construction of the trapping potentials mentioned in $({\bf I})$. To have an optical lattice that confines the fermions tightly and the bosons loosely, the energy difference between the ground state ($S$-state) and the  excited state ($P$-state) of fermions must be smaller than that of the bosons, $\Delta E_{f} < \Delta E_{b}$.  In this way, one can choose a laser with frequency $(\omega)$  red detuned with respect to both excitation energies, ($\hbar \omega <\Delta E_{f}, \Delta E_{b}$), such that the detuning for the fermions is smaller than that for the bosons,  $\Delta E_{b}-\hbar \omega > \Delta E_{f} - \hbar \omega$.  For $^{40}$K fermions, its 
$4S$ to $4P$ transition has a wavelength 740$nm$, where as the wavelength of the $2S$ to $2P$ transition of $^{7}$Li boson is 671$nm$, which satisfies the aforementioned condition. Moreover, the difference between these two excitation energies are large enough so that  the detuning $\Delta E_{b}-\hbar \omega$ and $ \Delta E_{f} - \hbar \omega$ sufficiently large to suppress heating due to spontaneous emission. Condition $({\bf I)}$ can therefore be satisfied.  

Another scheme that makes use of the hyperfine structure of the $P$ state of $^{87}$Rb was 
pointed out in ref.\cite{specific}.  By tuning the laser frequency to 790.01$nm$, which is between the two hyperfine states $P_{3/2}$ and $P_{1/2}$ of $^{87}$Rb,  Rb bosons sees no lattice because the red detuned lattice due to  $P_{3/2}$ is cancelled by the blue detuned lattice due to $P_{1/2}$.  On the other hand, such laser will generate an attractive potential for $^{6}$Li and $^{40}$K, since its wavelength is longer than those of the $S$-$P$ transitions of $^{6}$Li and $^{40}$K respectively.

To illustrate  our scheme, we shall first discuss the basic properties of lattice fermions and the important process of entropy-redistribution. 

\section{(B)  Number density and entropy distributions of lattice fermions:}
A Fermi gas in the lowest band of a (3D) optical lattice is described by the 
Hubbard model  
\begin{equation}
\hat{K} =  \hat{J}  + \hat{W}  - \mu \hat{N},
\label{Hubbard} \end{equation} 
where $\hat{J} = - J \sum_{\bf \langle R, R'\rangle, \sigma} a^{\dagger}_{\bf R, \sigma} a^{}_{\bf R', \sigma} $
describes hopping of fermions with spin $\sigma$, ($\sigma = \uparrow, \downarrow$) between neighbouring sites ${\bf R}$ and ${\bf R'}$, $J$ is the tunneling integral,   $\hat{W} = U\sum_{\bf R}^{}n^{}_{\bf R, \uparrow} n^{}_{\bf R, \downarrow}$
   describes the on-site repulsion $(U)$ between spin up and spin down fermions;   $a^{\dagger}_{\bf R, \sigma}$ and   $n^{}_{\bf R, \sigma} = a^{\dagger}_{\bf R, \sigma} a^{}_{\bf R, \sigma} $  are the creation and number operators of a fermion with spin $\sigma$  at site  ${\bf R}$; 
$\hat{N} = \sum_{\bf R, \sigma}n^{}_{\bf R, \sigma}$
 is the total fermion number,  and  $\mu$  is the chemical potential.  
 
The possible states of the fermions are: band insulator $(BI)$, Mott insulator $(MI)$, and "conducting" state $(C)$, corresponding to two, one and a non-integer number of fermions per site, respectively.  The Mott insulator will develop AF order at the Neel temperature $T_{N}\sim J^2/U$, while the conducting state is expected to have a superfluid ground state. 
The strongly correlated regime emerges when
\begin{equation}
U>>J >> J^2/U, 
\end{equation}
with $J^2/U$ being the smallest energy scale.  At present,  experiments operate in the temperature range 
\begin{equation}
U>T>J. 
\label{T-regime}\end{equation}
For simplicity, we set Boltzmann's constant $k_{B}=1$.  
 Our goal is to {\em perform operations in this high temperature regime so that the fermions will lose a substantial amount of entropy}. 

Within the temperature range described by eq.(\ref{T-regime}), $\hat{J}$ can be treated as a perturbation and $\hat{K}$ is site-diagonal to zeroth order in $J$. 
The number occupation at site ${\bf R}$ is then
 \begin{equation}
 n^{}_{\bf R}(T, \mu) = \sum_{\sigma} \langle \hat{n}^{}_{\bf R, \sigma}\rangle =  \frac{ 2e^{\mu/T} + 2 e^{(2\mu - U)/T}   }{ 1 + 2e^{\mu/T} + e^{(2\mu - U)/T}    } .
\label{nr} \end{equation}
The entropy per site is 
$s^{}_{\bf R} = \partial ( T{\rm ln}Z_{\bf R}^{})/\partial T $, where  
$Z_{\bf R}^{}(T,\mu) = {\rm Tr} e^{-\hat{K}_{\bf R}/T}
= 1 + 2e^{\mu/T} + e^{(2\mu - U)/T} $ is the partition function at site ${\bf R}$. Explicitly, we have 
\begin{equation}
s_{\bf R}^{}(T, \mu) =  {\rm ln}Z_{\bf R}^{}(T,\mu) + (E_{\bf R}^{}(T, \mu) -\mu n_{\bf R}^{}(T, \mu))/T
\label{sr} \end{equation}
and $E_{\bf R}^{}(T, \mu) = U\langle \hat{n}_{\bf R, \uparrow}^{} \hat{n}_{\bf R, \downarrow}^{}\rangle  = e^{(2\mu - U)/T}/Z_{\bf R}^{}(T,\mu)$.

In a harmonic trap $V_{\omega}^{}({\bf R}) = M\omega^2 R^2/2$, the  density and entropy distributions
in  the temperature range $U>T>J$ can be calculated from eq.(\ref{nr}) and eq.(\ref{sr}) using local density approximation (LDA) by replacing $\mu$ with $\mu({\bf R}) \equiv \mu - V_{\omega}^{}({\bf R})$. A typical distribution is shown in Figure 1. 
The chemical potential $\mu$ and temperature $T$ in eq.(\ref{nr}) and (\ref{sr}) are determined from the number and entropy constraints, 
 \begin{equation}
 N = \sum_{\bf R} n_{\bf R}^{}(T, \mu -  V^{}_{\omega}({\bf R})), \,\,\,\,\,\,  
  S = \sum_{\bf R}  s_{\bf R}^{}(T, \mu - V^{}_{\omega}({\bf R})),  
 \label{SNtotal}\end{equation}
where $N$ is the total number of fermions and $S$ is the total entropy produced  in the  convention cooling scheme before the lattice is turned on.  Figure 1 shows the following phases:   [$ (BI) : n_{\bf R}^{} \rightarrow 2$, for  $U< \mu({\bf R})$]; 
[$(MI) : n_{\bf R}^{} \rightarrow 1$ for  $0<\mu({\bf R}) < U$];  [({\em Vacuum}) : $n_{\bf R}^{} \rightarrow 0$ for    $ \mu({\bf R}) <0$]. 
There are also two  "conducting" phases $(C1)$ and $(C2)$ with non-integer number of fermions per site. They are 
$(C1):   1< n_{\bf R}^{} < 2$, with $\mu({\bf R}) \sim U$, and 
$ (C2):   0< n_{\bf R}^{} < 1$ with  $\mu({\bf R}) \sim 0$. 
We shall denote the centres of the conducting regions $(C1)$ and $(C2)$ as $R_{1}^{}$  and $R_{2}^{}$ . They are determined by 
\begin{equation}
\mu - \frac{1}{2}M\omega^2 R^{2}_{1}=U \,\,\,\,\, {\rm and} \,\,\,\,\, \mu - \frac{1}{2}M\omega^2 R^{2}_{2}=0. 
\label{R1R2} \end{equation}

\begin{figure}
\centerline{\includegraphics[width=.5\textwidth]{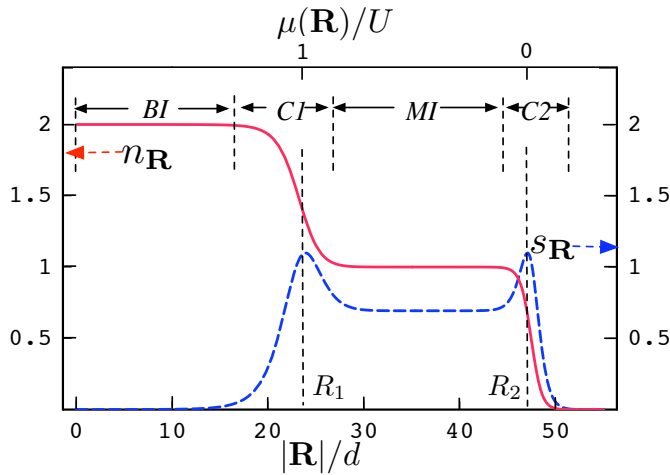}}
\caption{ Density distribution $n_{\textbf R}$ (red) and entropy distribution $s_{\textbf R}$ (blue) of a Fermi gas in the lowest band of an optical lattice and in the presence of an harmonic trap.  $BI$ and $MI$ denote band insulator and Mott insulator, respectively. The regions labeled $C1$ and $C2$ are the ``conducting" regions where the fermions have large number fluctuations and are mobile.  }\label{fig1}
\end{figure}

The $(BI)$ region has essentially zero entropy per particle, $s_{\bf R} \sim 0$, since the pair $(\uparrow\downarrow)$ is the only possible configuration at each lattice site.  
At  temperatures $U> T> T^{}_{N}$, $(MI)$ has spin entropy $s_{\bf R}\sim {\rm ln}2$, since both $\uparrow$ and $\downarrow$, are equally probable. At the same temperature range, $(C1)$ has  $s_{\bf R}\sim {\rm ln}3$,  since  the doublet $(\uparrow\downarrow)$ and the single spin states, $\uparrow$ and $\downarrow$ are all equally probable. At even higher temperatures, $s_{\bf R}$ of both $(MI)$ and $(C1)$ can rise as high as ${\rm ln}4$,  as the probability of having an empty site increases from zero. 
Similar situation occurs in $(C2)$.

While the total entropy and other properties of the system can be calculated in the temperature range $U>T>J$  using eqs.(\ref{nr}), (\ref{sr}), and (\ref{SNtotal}), it is useful to understand them using simple estimates. 
The total entropy of the system is 
\begin{equation}
S = 4\pi R^{2}_{1} \Delta R_{1} \frac{\overline{s}_{C}}{d^3} + 4\pi R^{2}_{2} \Delta R_{2} \frac{\overline{s}_{C}}{d^3} + 
\frac{4\pi}{3} (R^{3}_{2} -  R^{3}_{1}) \frac{\overline{s}_{M}}{d^3},
\end{equation}
 where  $d$  is the lattice spacing,   and $\Delta R_{1}$ and $\Delta R_{2}$   are the widths of the $(C1)$ and 
 $(C2)$ regions. They are given by $\Delta \mu(R_{1})\sim T$   and $\Delta \mu(R_{2})\sim T$, or   \begin{equation}
 \Delta R_1 = T/(M\omega^2 R_1) \,\,\,\,\, \Delta R_2 = T/(M\omega^2 R_2).
 \end{equation}
  The quantities   
 $\overline{s}_{C}$ and   $\overline{s}_{M}$  are the average entropy per site in $(C)$ and $(MI)$ regions:  ${\rm ln 3} <\overline{s}_{C}< {\rm ln}4$, and   ${\rm ln} 2< \overline{s}_{M} < {\rm ln 4} $. 
 From eq.(\ref{R1R2}), we also have 
 \begin{equation}
 R_{2}^2 - R^{2}_{1} = 2U/(M\omega^2),
 \end{equation}
 showing that the $(MI)$  region shrinks as $\omega$ increases. 
 
\section{(C) Entropy localization and reduction:} 
From Figure 1, we see that if we compress the trap {\em adiabatically} (by increasing $\omega$),  the band insulator will grow at the expense of other phases.
As a result, the Mott regions and the ``conducting" regions, and hence all the entropy, are pushed to the surface, as shown in Figures 2a and 2b.  Since all the entropy in the bulk is squeezed into the surface layer, the entropy density at the surface must rise above its former value to keep the total entropy constant, which means the temperature the system will increase, as shown in Figure 2. The
fact that adiabatic compression causes heating is a consequence of thermodynamics\cite{HoZhou}.
  
As compression proceeds, it reaches the point where the width of the $(MI)$ region becomes so thin that 
it is of the order of a few lattice spacings. (See Fig.2a and 2b). This occurs at $R_{2}-R_{1} << R_{1}\sim R_{2} \sim R$, where $R$  is the radius of the $(BI)$ region related to the total fermion number  $N$ by $N=2(4\pi/3)(R/d)^3$ . In this case, we have \begin{equation}
 \frac{S}{N} = \frac{3}{2} \left(  \frac{8\pi}{3N}\right)^{2/3}\left( \frac{2T\overline{s}_{C} + U \overline{s}_{M}}{M\omega^2 d^2}\right).
\label{SoverN} \end{equation} 			
Eq.(\ref{SoverN}) shows that $T$ rises as $\omega^2$ during an adiabatic compression. 
 
If, however, we are able to keep the temperature constant during the compression, the entropy density at the surface will remain at its initial value, (see Figure 2c, 2d). As a result, the total entropy will be reduced by the same amount as the reduction of entropic volume (i.e. the region that contains entropy), which is substantial as the latter changes from a 3D volume to a surface layer. 
This can also be seen in eq.(\ref{SoverN}), where $S/N$ drops  as $\omega^{-2}$ in an isothermal compression. 

 As isothermal compression continues, the widths of the $(C)$ and $(MI)$ regions will shrink down to a lattice spacing $d$,  and the entropic surface region will eventually be reduced to a surface of thickness $d$.  At this point, LDA breaks down. Due to thermal fluctuation, the thickness of this surface layer will remain approximately one lattice spacing, regardless of further increase of the trapping potential. As long as $T$  is above the spin ordering temperature  $T_{N}^{}$, the entropy per site is still ${\rm ln}2$. This limits the lowest entropy attainable in isothermal compression for  $T>J^2/U$ to  $S^{\ast}=(4\pi R^2 d)(\overline{s}/d^3)$, or 
 \begin{equation}
 \frac{S^{\ast}}{N} = \frac{3}{2} \left(  \frac{8\pi}{3N}\right)^{1/3}\overline{s}, 
\label{SNlimit} \end{equation}
where $\overline{s}\sim {\rm ln}2$. For $N =  5\times 10^6$ $(3\times 10^7)$, $S^{\ast}/N$ is 0.012  (0.007) which is 2$\%$ (1$\%$) of the lowest value attainable today.

\section{(D) Entropy transfer between fermions and BEC:}  
To keep the fermions at roughly constant temperature, they must be in contact with a heat reservoir. A BEC is an ideal medium for this purpose, for it has a much higher heat capacity than the fermions and can therefore keep the temperature roughly constant.  In this section, we shall provide explicit 
calculations to demonstrate this fact.  Our results show that the transfer of entropy  between fermions and bosons is accurately described by the simple isothermal compression model in Section ${\bf (C)}$.

We shall consider a mixture of BEC and fermions. The BEC is contained in a harmonic trap  $V^{}_{B}({\bf R}) = \frac{1}{2}M^{}_{B} \omega_{B}^{2} {\bf R}^2$ with frequency $\omega_{B}$, where $M_{B}$ is the boson mass, and another trap 
$V^{}_{F}({\bf R}) = \frac{1}{2}M^{}_{F} \omega_{F}^{2} {\bf R}^2$ is used 
for the fermions.  The fermions are also confined in an optical lattice. We shall assume these potentials are species specific as described in summary ${\bf (I)}$ in Section $({\bf A)}$, so that $V^{}_{F}$ can be varied independently of $V^{}_{B}$. This is important because the compression of fermions should not lead to a substantial compression of the BEC. Otherwise the temperature of the BEC will rise, making it less efficient in absorbing the entropy of the fermions. 

To study the entropy transfer between fermions and bosons, we first 
consider a homogenous Bose-Fermi system.  In the grand canonical ensemble, the hamiltonian is
\begin{equation}
\hat{K}=\hat{K}_{F}(\mu_{F})+\hat{K}_{B}(\mu_{B}) + \hat{H}_{BF}, 
\end{equation}
where $\hat{K}_{F}$ is the Hubbard hamiltonian eq.(\ref{Hubbard}) with $\mu$ now denoted as $\mu_{F}$.   $\hat{K}_{B}$ is the hamiltonian for bosons  with chemical potential $\mu_{B}$,
\begin{equation}
K_{B} = \int \left( \frac{\hbar^2}{2M_{B} } \nabla  \hat{\phi}^{\dagger} \cdot \nabla \hat{\phi}
+ \frac{g_{BB}}{2}  \hat{\phi}^{\dagger} \hat{\phi}^{\dagger} \hat{\phi}^{}  \hat{\phi}^{}-\mu_B\hat{\phi}^{\dagger}  \hat{\phi}\right), 
\end{equation}
where $ \hat{\phi}^{\dagger} $ is the creation operator for bosons and $g_{BB} = 4\pi\hbar^2 a_{BB}/M_{B}$.  $\hat{H}_{BF}$ is the boson-fermion interaction, 
\begin{equation}
H^{}_{BF} = \frac{g^{}_{BF}}{d^3} \sum_{{\bf R}} \hat{n}^{}_{F, {\bf R}} \hat{n}^{}_{B, {\bf R}}, 
\end{equation}
with coupling constant $g_{BF}^{}$, $d$ is the lattice spacing, and $\hat{n}^{}_{F, {\bf R}}$ ($\hat{n}^{}_{B, {\bf R}}$) is the number of fermions (bosons) in a unit cell centered at site ${\bf R}$. 

\begin{figure}[tp]
\centerline{\includegraphics[width=.5\textwidth]{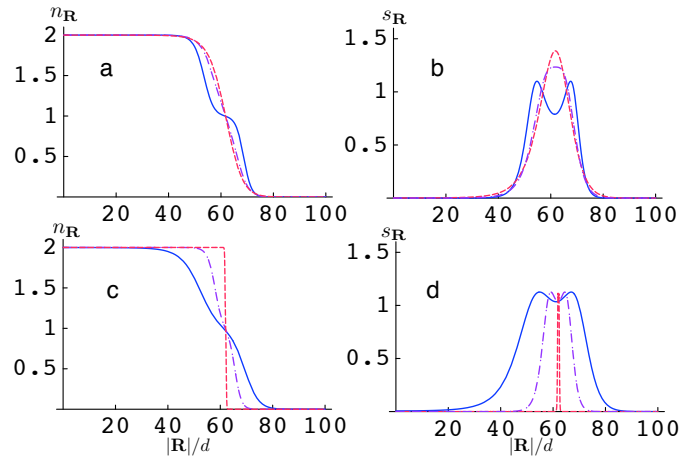}}
\caption{Figure (2a) shows the number density $n_{\textbf R}$ of $N=2 \times 10^{6}$  $^{40}$K fermions  in an optical lattice with lattice height $V_{o}=15E_{R}$ and $U=0.5E_{R}$ at  harmonic trap frequencies $\nu_{F}=\omega_{F}/2\pi=$ 100Hz (blue), 150Hz (purple) and 600Hz (red),  calculated from eqs.(\ref{nr})-(\ref{SNtotal}).  The entropy per particle is fixed at $S/N=0.5$.
 The corresponding entropy distributions $s_{\textbf R}$ are shown in Figure (2b), with temperatures 
27nK (blue), 88nK (purple), and 1880nK (red), respectively. 
As the trap frequency is increased from 100Hz (blue)  to 150Hz (purple), Figs.(2a) and (2b) show that  the Mott phase melts away and the entropy density at the center of the conducting layer increases, leading to a rise in temperature. This rise is very rapid.  It scales as $\omega^2$ as seen in eq.(\ref{SoverN}).  Figures 
(2c) and (2d) show the number density $n_{\textbf R}$ and entropy density of $s_{\textbf R}$ for the same parameters as Figure (2a) and (2b), but now with the temperature fixed ($T=50$nK) instead of entropy per particle. 
As the harmonic trap is compressed from 100Hz (blue) to 150Hz (purple) and then 600Hz (red), $S/N$  decreases from 0.732 (blue) to 0.333 (purple) and finally, 0.021 (red), reaching a value which is only 4$\%$  of the lowest value attainable today, and has not yet reached the limit shown in eq.(\ref{SNlimit}). 
%As the trap frequency increases to 150Hz (purple), Fig.(2a) and (2b) show that  the Mott phase has melted away, and the entropy density at the center of the conducting layer increases, leading to a temperature rise. This rise is very rapid.  It scales as $\omega^2$ as seen in eq.(\ref{SoverN}). Figure  (2c) and (2d) show the number density $n_{\textbf R}$ and entropy density of $s_{\textbf R}$ of the fermion system in figure 2a, now kept at a constant temperature, $T_{i}=50$nK.   As the harmonic trap is compressed from 100Hz (blue), 150Hz (purple), to 600Hz (red), $S/N$  decreases as  0.732 (blue), 0.333 (purple), and 0.021 (red), reaching a value which is only 4$\%$  of the lowest value attainable today.
}\label{fig2}
\end{figure}

To calculate the properties the system, we make the following approximations : 

\noindent (i) a mean field decomposition of $H_{BF}$, 
replacing it by $H_{BF}^{M} = g_{BF}\sum^{}_{{\bf R}} ( \hat{n}^{}_{F, {\bf R}} n^{}_{B}  + 
n^{}_{F} \hat{n}^{}_{B, {\bf R}} - n^{}_{F} n^{}_{B} )$, where $n_{F}^{}=\langle \hat{n}^{}_{F, {\bf R}}\rangle$; 
 $n_{B}^{}=\langle \hat{n}^{}_{B, {\bf R}}\rangle$; 

\noindent  (ii) treating the hopping term $J$ as a perturbation of $\hat{K}^{}_{F}$ as in Section {\bf (B)}.  This is justified in the temperature regime $U>T>J$; 

\noindent (iii) applying the Hartree-Fock approximation for thermodynamics of the bosons\cite{Stringari}. With  approximation (i), $\hat{K}$ becomes 
\begin{equation}
\hat{K}'= \hat{K}^{}_{F}(\mu^{}_{F} - g^{}_{BF} n_{B}) +  \hat{K}^{}_{B}(\mu^{}_{B} - g^{}_{BF} n^{}_{F} )
- g^{}_{BF} n^{}_{B} n^{}_{F}. 
\label{K'} \end{equation}
The pressure $P(T, \mu_{F}, \mu_{B})= \Omega^{-1}T{\rm ln} {\rm Tr} e^{-K'/T}$ is then 
\begin{eqnarray}
P(T, \mu_{B}, \mu_{F}) & = P^{}_{F}(T, \mu^{}_{F} - g^{}_{BF} n^{}_{B}) \hspace{0.9in} \nonumber \\
  &+ P^{}_{B}(T, \mu^{}_{B} - g^{}_{BF} n^{}_{F}) + g_{BF}^{}n_{F}^{}n_{B}^{}, 
\end{eqnarray}
where  $P^{}_{F,B}(T, \mu^{}_{F,B})= \Omega^{-1} T{\rm ln}e^{-K^{}_{F,B}(\mu^{}_{F,B})/T}$, and $\Omega$ is the volume of the system.  The fermion and boson densities are given by 
 \begin{equation}
n_{F} = \frac{\partial P}{\partial \mu_{F}},  \,\,\,\,\,\,\,  n_{B} = \frac{\partial P}{\partial \mu_{B}}.
\end{equation}
Within approximation (ii), the fermion density is readily given by eq.(\ref{nr}) as 
\begin{equation}
n_{F} = n^{}_{F}(T, \mu^{}_{F} - g^{}_{BF} n^{}_{B}).
\label{nF} \end{equation}
To apply approximation (iii), we follow the procedure in reference \cite{Stringari} to calculate the boson density for {\em both} the normal and superfluid parts of the boson cloud, which is of the form
\begin{equation}
n_{B} = n^{}_{B}(T, \mu^{}_{B} - g^{}_{BF} n^{}_{F})
\label{nB} \end{equation}
 according to eq.(\ref{K'}). (The presence of the superfluid will depend on the value $ \mu^{}_{B} - g^{}_{BF} n^{}_{F}$). 
 
Eqs.(\ref{nF}) to (\ref{nB}) form a complete set of equations that determine ($n^{}_{B}$, $n^{}_{F}$) self consistently as a function of $(T, \mu_{B}^{}, \mu_{F}^{})$.   All these have to be  evaluated numerically. 
 These solutions then allow one to evaluate the entropy density, which is 
 $s(T, \mu_{F}^{}, \mu_{B}^{})  = \frac{\partial P}{\partial T}$, or 
\begin{equation}
s =  \frac{\partial P}{\partial T} = s_{F}^{} + s^{}_{B} + s^{}_{BF},
\end{equation}
where $s_{F}^{}, s_{B}^{}, s_{BF}^{}$ are defined as fermion, boson, and ``interaction" entropy density respectively, 
\begin{equation}
s_{F}^{} \equiv  \frac{\partial P^{}_{F}}{\partial T}, \,\,\,\,\, s_{B}^{} \equiv  \frac{\partial P^{}_{B}}{\partial T}, \,\,\,\,\,
s_{BF}^{} \equiv  g_{BF}^{} \frac{ \partial (n_{B}^{} n_{F}^{}) }{\partial T},
\end{equation}
which will also be functions $(T, \mu_{B}^{}, \mu_{F}^{})$. To find the density and entropy distribution in the presence of the boson and fermion traps, we can use LDA to replace $\mu^{}_{F,B}$ by  $\mu^{}_{B,F}({\bf R}) = \mu_{B,F}^{} - V_{B,F}^{}$ . 

\begin{figure}[tp]
\centerline{\includegraphics[width=.5\textwidth]{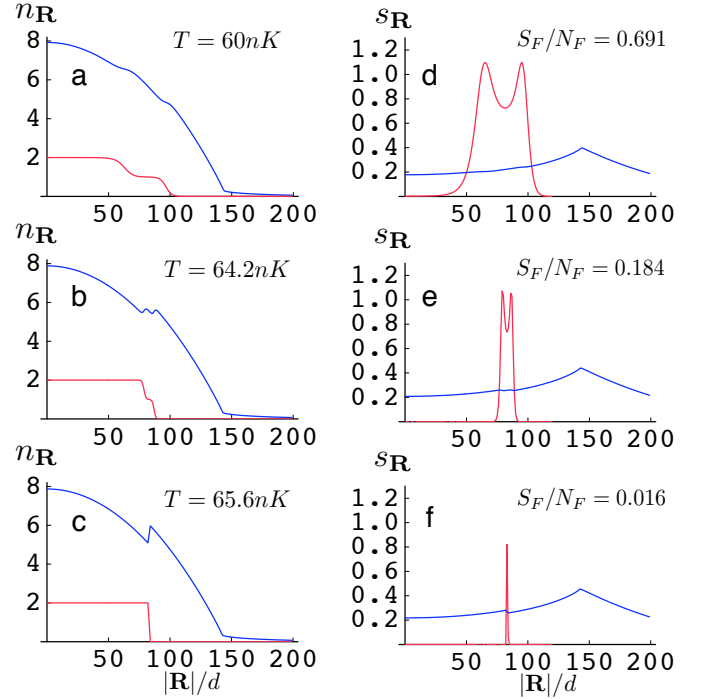}}
\caption{Density distribution $n_{\bf R}^{}$ and entropy distribution $s_{\bf R}^{}$ of bosons (blue) and fermions (red) in a mixture of BEC and lattice Fermi gas. The total number of fermions $N_F$ (Bosons $N_B$) is $5\times10^6$ ($5\times10^7$). The trapping frequency of bosons is fixed at $\nu_B=10 Hz$.
The number density of bosons and fermions at fermion harmonic trap frequencies  $100Hz$, $200Hz$, and $600Hz$ are shown in Figs.(3a)-(3c). The corresponding entropy densities are shown in Figs. (3d) to (3f).
As the frequencies  $\nu_F$ of the fermion harmonic trap increases, the edge of the band insulator sharpens  and the entropy distribution becomes more and more narrowed, yet
the temperature of the system changes very little as the fermion trap is tightened, from $60nK$ to $64.2nK$, and to $65.6nK$. As a result, the  height of the entropy distribution remains roughly constant. The narrowing of the entropy distribution, however, leads to a rapid drop in the entropy per particle, from 0.691  to 0.184, to 0.016, as shown in Figs.(3d)-(3f).  }\label{fig4}
\end{figure}

In our numerical calculations, we consider $5\times 10^{6}$ of $^{40}$K fermions in a lattice immersed in a BEC of $^{87}$Rb with $N^{}_{B}=5\times 10^7$ bosons.  The frequency of the fermion trap is initially set at $\nu^{}_{F}= \omega^{}/(2\pi)$ = 100Hz. The bosons are confined in a loose trap with frequency $\nu^{}_{B}= \omega^{}_{B}/(2\pi) = 10$Hz, so that it covers the entire fermion system. 
We take $a_{BB}^{}=5.45$ nm, and $g_{BF}^{}= g^{}_{BB}/2$.   The lattice height is $V_{o}=15E_{R}$ and the fermions have a Hubbard interaction  $U=1.5E_{R}$, where $E_{R}$ is the recoil energy. To demonstrate the transfer of entropy from the fermions  to the BEC, we consider an initial state with initial temperature 60nK, corresponding to the entropy per particle  $S_{F}/N_{F}=0.691$ for the fermions.  For these parameters, most of the bulk is already a band insulator, similar to that in Figure (2a) and the entropy density is accumulated at the surface. (See Fig.(3a)). 

When the fermion trap frequency $\nu^{}_{F}$ increases from 100Hz to 600Hz, we see from Figs. (3a)-(3c)
that the edge of the band insulator becomes sharper, while the entropy density becomes concentrated
at the surface (Fig. (3d)-(3f)).   Our calculation also show that the fermion entropy $S^{}_{F} \equiv \int s^{}_{f}$ decreases whereas the boson entropy $S_{B}^{}=\int s_{B}^{}$ increases, while $S_{BF }\equiv \int s_{BF}^{}$ remains much smaller than $S_{F}$ and $S_{B}$ during the compression, and the entropy of the entire Bose-Fermi mixture is a constant.  

Due to the large heat capacity of the bosons,  
the overall temperature $T$ rises only moderately, from 60nK to 65.6 nK.  This shows that 
the simple isothermal compression model for the lattice fermions in Section $({\bf C})$ is a reasonable approximation of the entropy transfer process between the bosons and the fermions. At $\nu^{}_{F}$=600Hz, the fermion entropy per particle is $S^{}_{F}/N^{}_{F}\sim 0.016$, which is about  $3\%$ of the best estimate achievable with conventional methods. 

\section{(E):  Equilibration of the lattice fermions after boson evaporation:}
%After transferring the entropy of the fermions to the BEC at temperature $T^{(i)}$  by tightening the trap to $\nu^{}_{F}$ in the Bose-Fermi mixture discussed in Section ${\bf (D)}$,  ($T^{(i)}\sim 65$nK and $\nu^{}_{F} = 600$Hz), 
After transferring the entropy of the fermions to the BEC as shown in Fig.(3c) at the temperature $65.6\,{\rm nK}(\equiv T^{(i)})$ and at trap frequency $\nu^{}_{F} = 600$Hz,  
 we evaporate all the bosons suddenly so as to obtain a pure fermion system. 
The lattice Fermi gas will then relax to its equilibrium state corresponding to the new trap frequency  $\omega^{}_{F}= 2\pi \nu^{}_{F}$. In this process, entropy will be generated and temperature will rise. However, because the band insulator is incompressible, and because the trap has a larger value of $\omega^{}_{F}$, entropy generation is limited. We find (in the calculation below) that the increases in entropy and temperature are about a few percent of their values prior to evaporation.  
In other words, the entropy of the band insulator after it reaches equilibrium is 
essentially the same as that before boson evaporation, i.e. $S_{F}^{}/N_{F}\sim 0.02$.  This will be the entropy inherited by the Mott insulator that emerges from the band insulator as one decompresses the trap.  

The calculation for the above processes is as follows. 
Immediately after evaporation, the lattice  Fermi gas  has total energy 
\begin{eqnarray}
E^{(i)}= & \sum_{\bf R}  \epsilon (T^{(i)}, \mu_{F}^{} - V_{\omega}({\bf R}) - g^{}_{BF} n_{B}({\bf R}) )  \nonumber \\
 & + \sum_{\bf R}V_{\omega}^{}({\bf R}) n_{F}^{}({\bf R}), 
\end{eqnarray} 
where
$\epsilon(T, \mu) \equiv  U \langle \hat{n}_{{\bf R}, \uparrow} \hat{n}_{{\bf R}, \downarrow}\rangle$
is the internal energy per site of a {\em homogeneous} lattice fermion with hamiltonian $\hat{K}_{F}^{}$ at the temperature range $U>T>J$, and 
\begin{equation}
\epsilon (T, \mu)= \frac{Ue^{(2\mu-U)/T}}  {1 + e^{\mu/T} + e^{(2\mu-U)/T}}. 
\end{equation}
$n_{B,F}^{}({\bf R})$ are boson and fermion densities prior to boson evaporation, which were calculated from eqs.(\ref{nF}) and (\ref{nB}).   When the system finally relaxes to equilibrium, it will have a different temperature $T^{(f)}$ and chemical potential $\mu_{F}^{(f)}$. The energy of the final system is 
\begin{eqnarray}
E^{(f)}= & \sum_{\bf R}  \epsilon (T^{(f)}, \mu_{F}^{(f)} - V_{\omega}({\bf R}) ) \nonumber \\
 & + \sum_{\bf R}V_{\omega}^{}({\bf R}) n_{F}^{} (T^{(f)}, \mu_{F}^{(f)} - V_{\omega}({\bf R}). 
\end{eqnarray} 
Since energy is conserved in this process, we have $E^{(f)}= E^{(i)}$. In addition, the total number of particles is given by $N_{F} = \sum_{\bf R}  n_{F}^{} (T^{(f)}, \mu_{F}^{(f)} - V_{\omega}({\bf R}) )$. These two relations uniquely determine the final temperature $T^{(f)}$ and chemical potential $\mu_{F}^{}$ of the band insulator, from which one can calculate the final entropy.  With the parameters we mentioned before, we find that the entropy increase in this process is negligible, only about a few percent of the value before evaporation;  hence our conclusions summarized in Section $({\bf A})$.  

\section{Final Remarks:} 
We have introduced a method that allows extraction of a substantial fraction of  the entropy of a Fermi gas in an optical lattice after a strong lattice is switched on. 
Moreover, the extraction process is  conducted at the temperature regime $T>J$, much higher than the Neel temperature $T_{N}\sim J^2/U$. 
{\em While our method makes explicit use of the band insulator, it is applicable to any system which has an equilibrium phase with a large gap. } The idea is to use the gapful phase to push away all the entropy in the bulk into a surrounding Bose-Einstein condensed gas.
It should also be stressed that although the final stage of our method involves 
evaporating away the Bose-Einstein condensed gas, it  is different from the usual sympathetic cooling both in purpose and in function. First, our evaporation must be preceded by the localization of fermion entropy in order to be effective. Second,  the purpose of removing the bosons is not to decrease the energy of the system (as in the usual sympathetic cooling), but to remove the entropy it absorbed from the fermions. 
A natural question is whether one can efficiently reduce the entropy of a Mott phase which is in direct contact with a BEC by evaporating on the latter.  The answer is negative, as shown by our calculations. The reason is that the interactions between bosons and fermions typically have very weak spin dependence. The removal of bosons therefore has little effect on the spin entropy of the Mott phase, which will remain close to its value before boson evaporation, ${\rm ln}2$ per particle. 

Finally, we point out that our scheme assumes that it is possible to expand the trap adiabatically to turn a band insulator into a Mott insulator. This implies that the relaxation time for particle re-distribution is sufficiently fast. 
Since mass transport occurs at the interface between different phases, (i.e. regions $(C1)$ and $(C2)$),  it will take place within the time scale $\hbar/J$. It is therefore helpful to use lattices with sufficiently large tunneling, (say, around $10-15E_{R}$), and to reach the large U limit  by increasing the interaction using a Feshbach resonance\cite{Chin}.

This work is supported by NSF Grants DMR0705989, PHY05555576, and by DARPA under the Army Research Office Grant No. W911NF-07-1-0464.  We thank Randy Hulet for discussions on creating different potentials for bosons and fermions, and Ed Taylor for a careful reading of the manuscript.


\begin{thebibliography}{99}
\bibitem{Cho} Cho A (2008), CONDENSED-MATTER PHYSICS: The Mad Dash to Make Light Crystals, \textit{Science} 320: 312.
%2. Chin, J.K. et al. Evidence for Superfluidity of Ultracold Fermions in an Optical Lattice. Nature. 443, 961-964 (2006).
%3. Cirac, J.I. & Zoller, P. How to Manipulate Cold Atoms. Science. 301, 176-177 (2003). 
%4. Duan, L.M., Demler, E. & Lukin, M.D. Phys. Rev. Lett. 91, 090402 (2003).
\bibitem{HoZhou}  Ho TL,  Zhou Q (2007), Intrinsic Heating and Cooling in Adiabatic Processes for Bosons in Optical Lattices, \textit{Phys Rev Lett} 99: 120404.
\bibitem{adiabatic} Greiner M, Mandel O, Esslinger T, H\"ansch TW, Bloch I  (2002), Quantum phase transition from a superfluid to a Mott insulator in a gas of ultracold atoms,  \textit{Nature} 415: 39-44. 
\bibitem{Kohl} K\"ohl, M, Moritz H, St\"oferle T, G\"unter K, Esslinger T (2005), Fermionic Atoms in a Three Dimensional Optical Lattice: Observing Fermi Surfaces, Dynamics, and Interactions, \textit{Phys Rev Lett} 94: 080403.
\bibitem{AG} Werner F, Parcollet O, Georges A, Hassan SR (2005), Interaction-Induced Adiabatic Cooling and Antiferromagnetism of Cold Fermions in Optical Lattices,  \textit{Phys Rev Lett} 95: 056401.
\bibitem{Tremblay} (2007) Dar\'e A.-M., Raymond L, Albinet G, Tremblay A.-M. S \textit{Phys Rev B} 76: 064402.
\bibitem{Castin} Carr LD, Shlyapnikov GV, Castin Y (2004) \textit{Phys Rev Lett} 92: 150404.
\bibitem{specific} LeBlanc L, Thywissen J (2007), Species-specific optical lattices,  \textit{Phys Rev A} 75: 053612.
\bibitem{Stringari}Pitaevskii L, Stringari  S (2003) Bose-Einstein Condensation(Oxford University Press, Oxford), chapter 13. 
\bibitem{Chin}In the case of bosons, Cheng Chin's group at Chicago has recently succeeded in producing a Mott phase of Cesium bosons using a Feshbach resonance in a lattice with relatively large tunneling, which has shortened considerably the equilibration time. 
\end{thebibliography}
\end{document}